\documentclass{iaus}
\usepackage{graphicx}

\title[Spectral evolution models] 
{Spectral Evolution Models\\ for the Next Decade}

\author[Claus Leitherer]   
{Claus Leitherer}

\affiliation{Space Telescope Science Institute, 3700 San Martin Dr., Baltimore, MD 21218, USA
\\email: {\tt leitherer@stsci.edu}}

\pubyear{2008}
\volume{262}  
\pagerange{1--8}
\jname{Stellar Populations -- Planning for the Next Decade}
\editors{G. Bruzual \& S. Charlot}
\begin{document}

\maketitle

\begin{abstract}
Spectral evolution models are a widely used tool for determining the stellar content of galaxies. 
I provide a review of the latest developments in stellar atmosphere and evolution models, with an
emphasis on massive stars. In contrast to the situation for low- and intermediate-mass stars,
the current main challenge for spectral synthesis models are the uncertainties and rapid revision
of current stellar evolution models. Spectral libraries, in particular those drawn from theoretical 
model atmospheres for hot stars, are relatively mature and can complement empirical templates for
larger parameter space coverage. I introduce a new ultraviolet spectral library based on theoretical 
radiation-hydrodynamic atmospheres for hot massive stars. Application of this library to 
star-forming galaxies at high redshift, i.e., Lyman-break galaxies, will provide new insights into
the abundances, initial mass function and ages of stars in the very early universe.  
\end{abstract}

\firstsection 
\section{Introduction}

Spectral evolution models attempt to reproduce the observed spectra of systems ranging in sizes from star clusters to luminous galaxies by numerically 
combining models or observations of star formation, stellar evolution, and stellar spectra. \cite[Tinsley (1968)]{Tinsley68} is generally credited for initiating the field,
and \cite[Charlot \& Bruzual's (1991)]{ChaBru91} introduction of the isochrone synthesis technique validated this approach for ages up to a Hubble time. The success of synthetic spectral models echoes
the increased availability of computers in astronomy. The power of this technique has been amplified over the past decade by the growing importance of the internet and databases for 
storage, query, and distribution of spectral templates and related models. This trend is revolutionizing the field in a way similar to Tinsley's and
Charlot's \& Bruzual's achievements.
 
In this talk I will concentrate less on the technical but more on some of the astrophysical aspects of spectral evolution models. After a brief discussion of the main
ingredients in the model I will address the two major components: stellar evolution models and stellar libraries. The emphasis will be on {\em massive stars}, for which
stellar evolution still is the major challenge, whereas reliable stellar libraries are becoming available in larger and larger numbers. This situation is somewhat 
opposite to the case of low-mass stars. I will introduce a new ultraviolet (UV) spectral library constructed from a grid of radiation-hydrodynamic models and discuss its
potential for interpreting the rest-frame UV spectra of Lyman-break galaxies. Detecting the first stellar generations and modeling and understanding their spectra will
be a major effort for the next decade.

\section{Main Physics Input}

Evolutionary spectral synthesis is a special case of spectral synthesis whose goal is to reproduce observed spectra self-consistently from the star-formation history
of a galaxy and from stellar evolution models. The advantage of this method compared with spectral synthesis are the smaller number of free parameters and the increased
predictive power. The obvious price to pay is the reliance on models and idealized empirical parameters whose failure may doom the calculated synthetic spectra
(\cite[Cid Fernandes et al. 2005]{Cid05}). Since I am focusing on evolutionary synthesis in the following, assessing these assumptions becomes critical.

The astrophysical ingredients entering evolutionary spectral synthesis fall into four categories: (i) Quantities related to the star-formation process, i.e.,
the star-formation rate and its evolution with time and the stellar mass spectrum at  birth, also known as the initial mass function (IMF). Both are usually considered
free parameters and will not be discussed here. See, e.g., the conference proceedings of \cite[Corbelli et al. (2005)]{Cor05} and the review by \cite[Kroupa (2007)]{Kro07}. (ii) Once stars have formed, a prescription for the evolution of luminosity 
($L$), effective temperature ($T_{\rm eff}$), and mass ($M$) as a function of time and initial mass and chemical composition is needed. This prescription is provided by
stellar evolution models. (iii) Spectral libraries describe the emergent spectrum of each star for any given ($L$, $T_{\rm eff}$, $M$). These libraries can be either 
empirical or theoretical. (iv) Second-order effects, such as dust reddening or geometric effects are often accounted in  a very approximate way or even neglected altogether. 
Reddening by dust is of course a major issue when dealing with the UV. See the review by \cite[Calzetti (2001)]{Cal01}.

The areas of massive star evolution and stellar libraries are particularly active, with major revisions and anticipated new breakthroughs. I will therefore concentrate on these
two subjects in what follows.

\section{Stellar Evolution}

Until very recently the evolution of massive stars was thought to be determined by the chemical composition,
stellar mass, and mass-loss rate, plus atomic physics and some second-order effects. The evolution models
suggested reasonable agreement with observations of both individual stars and stellar populations. Subsequently the key role of stellar rotation in the
evolution of massive stars was recognized (\cite[Meynet 2009]{Mey09}). Evidence of anomalous stellar surface abundances already on the main-sequence, 
the lifetimes of certain evolutionary phases, and revised lower mass-loss rates support the concept of rotation.

\begin{figure}[t]
\begin{center}
 \includegraphics[width=3.0in]{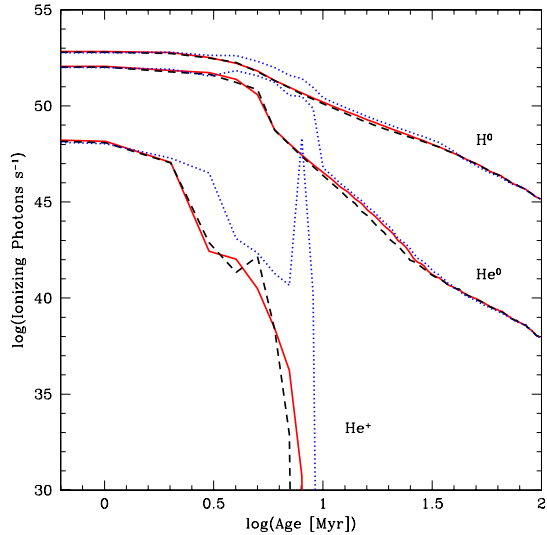} 
 \caption{Number of photons in the H$^0$, He$^0$, and He$^+$ continuum (solar composition). Models without  $v_{\rm rot} = 300$~km~s$^{-1}$ (solid) 
and with  $v_{\rm rot} = 0$ (dotted). The dotted curve denotes an alternative exploratory model
with $v_{\rm rot} = 0$. See \cite[V\'azquez et al. (2007)]{Vaz07} for details.}
   \label{fig1}
\end{center}
\end{figure}

Rotation modifies the hydrostatic structure, induces additional mixing and affects the stellar mass loss. Generally, rotation increases both
$L$ and $T_{\rm eff}$ in massive stars. This is the result of
the larger convective core and the lower surface opacity for higher rotation speed. (Recall that hydrogen is
the major opacity source and any decrease of its relative abundance by mixing lowers the opacity and therefore increases
the temperature.) The most dramatic changes with respect to models without rotation occur at the short-wavelength end of the spectrum. The ionizing
luminosities for a single stellar population with mass $10^6$~$M_\odot$ are shown in Fig.~\ref{fig1}. Since the most massive stars
are more luminous and hotter, their ionizing luminosities increase during O-star dominated phases (2~--~10~Myr).
The increase reaches a factor of 3 in the hydrogen ionizing continuum and several orders of magnitude in 
the neutral and ionized helium continua. The predictions for the latter need careful scrutiny, as the photon
escape fraction crucially depends on the interplay between the stellar parameters supplied by the evolution
models and the radiation-hydrodynamics of the atmospheres. In contrast, the escape of the hydrogen ionizing
photons has little dependence on the particulars of the atmospheres and consequently is a relatively safe 
prediction --- {\em if the stellar evolution models themselves can be trusted}.

Stellar winds from hot stars are a second area in which major changes have occurred. If our basic understanding of the evolutionary connections between 
different stellar species is correct, strong mass loss must occur in order to decrease the mass of an initially massive O star to that of a 
much less massive Wolf-Rayet star or red supergiant. Until recently, observational mass-loss rates were thought to be sufficiently well understood that 
they could serve as a fixed input for evolution models (\cite[Maeder \& Conti 1994]{Mae94}). The discovery of wind inhomogeneities with the result of much lower
mass-loss rates (\cite[Fullerton et al. 2006]{Ful06}; \cite[Puls et al. 2008]{Pul08}) and of eruptive phases (\cite[Smith 2010]{Smi10}) have dramatically changed this perspective. 
The uncertainties of the observed rates can easily reach factors of several. Therefore mass loss in evolution models is now considered an 
adjustable parameter that together with rotation governs the stellar properties with time.

The downward revision of the mass-loss rates of unevolved and evolved luminous, massive stars poses a severe challenge for the standard stellar evolution paradigm. 
Consider a 40~$M_\odot$ O star on the main-sequence, which is thought to become a Wolf-Rayet star on a time-scale of $\sim$5 Myr. Assuming a typical Wolf-Rayet mass of 15~$M_\odot$, the 
average mass-loss rate prior to entering the Wolf-Rayet phase should be $\sim$10$^{-5}$~$M_\odot$~yr$^{-1}$, almost an order of magnitude higher than the observed O-star rates. 
\cite[Smith (2010)]{Smi10} emphasized the potential role of eruptive phases of massive stars to compensate for the lower steady-state rates. During a short evolutionary phase massive stars could lose as much mass as during the entire O-star phase with a steady wind. 

Both the effects of rotation and the uncertain mass-loss rates make stellar evolution models of massive stars much less certain than previously thought, in particular for post-main-sequence
evolution. Therefore spectral evolution models relying on such phases must be viewed with care.

\section{Spectral Libraries}

Spectral libraries can be built from empirical or theoretical stellar spectra. The pros and cons of either approach are discussed extensively in the literature
(e.g., \cite[Gustaffson et al. 2007]{Gus07}; \cite[Martins \& Coelho 2007]{Mar07}).\\
 
Arguments in favor of empirical libraries are:
\begin{itemize}
\item
Laboratory data are often lacking and therefore the quality and completeness of the line lists used for the computation of stellar
atmospheres are insufficient.
\item
The computational effort for generating a large grid of models can be formidable. Even if individual models matching the observations can be produced, they
are not suitable for inclusion in a library if the required parameter space cannot be covered.
\item
Departures from local thermal equilibrium (LTE) can be significant and greatly complicate the computational effort.
\item
The most luminous stars have stellar winds and are extended. Therefore model atmospheres must account for sphericity effects and often include hydrodynamics.
\item
Deviations from spherical symmetry can be important, as well as depth effects in the photosphere. As a result, plane-parallel, one-dimensional models are no longer valid.\\
\end{itemize}

On the other hand, limitations of empirical libraries are:
\begin{itemize}
\item
Massive stars are rare. Assembling a complete library, in particular at non-solar chemical composition, can be a severe challenge.
\item
Telescope time is expensive, whereas computer time is cheap. Using a telescope for building a large stellar library is often impractical.
\item
Interstellar reddening by dust for massive stars is not negligible. This is particular true in the UV, where large reddening corrections must be
applied to observations.
\item
Related to the previous item are interstellar absorption lines in the UV. Such lines are blended with stellar lines and are almost impossible to correct for.
\item
An often overlooked issue is the need of a spectral-type vs. $T_{\rm eff}$ relation when empirical libraries are linked to evolution models. This relation is 
derived from models, so that ultimately empirical libraries depend on the same atmosphere models they are intended to replace.
\end{itemize} 

When assessing the relevance of the above points for different classes of massive stars, it turns out that concerns about theoretical stellar spectra mostly
apply to cool stars. The low temperatures of late-type stars magnify the challenges for atmosphere modeling so that cool, massive stars should be 
represented by empirical spectra in spectral evolution models. The opposite can be said for hot stars. Here, the atmosphere models are rather mature
(\cite[Puls 2008]{Puls08}) and any remaining shortcomings by far outweigh the drawbacks of observational libraries, such as parameter space coverage and 
contamination by interstellar gas and dust.

Following this philosophy, we have been pushing for a fully theoretical UV stellar library for implementation in the Starburst99 synthesis code (\cite[Leitherer et al. 1999]{Lei99};
\cite[V\'azquez \& Leitherer 2005]{Vaz05}; \cite[Leitherer \& Chen 2009]{Lei09}). The main science driver are the rest-frame UV spectra of Lyman-break galaxies whose
spectral features may hold the clue for understanding star formation at redshifts of 3 and higher. Lyman-break galaxies are massive ($M \approx 10^{11}~M_\odot$), mildly
metal-poor (O/H+12~$\approx 8.2$), UV-bright ($E(B-V) \approx 0.1$) galaxies currently forming stars at rates of order $10^2~M_\odot$~yr$^{-1}$ (\cite[Giavalisco 2002]{Gia02}).
Prior attempts to model their rest-frame UV spectra using an empirical library were quite successful (e.g., \cite[Quider et al. 2009]{Qui09}) but the limitations are
obvious: the lack of spectra of metal-poor massive stars, the presence of $\alpha$-element/Fe variations, and the generally low S/N of the UV spectra of the template stars.
A first effort to model Lyman-break rest-frame UV spectra purely theoretically was done by \cite[Rix et al. (2004)]{Rix04}, who matched the weak photospheric features in the spectra to
determine stellar chemical abundances. The state of the stellar atmospheres at that time precluded the modeling of the stellar-wind lines in the spectra, which is 
frustrating because the wind lines are the strongest spectral features and can be detected even in low-S/N spectra of Lyman-break galaxies.

\begin{figure}[t]
\begin{center}
 \includegraphics[width=3.4in,angle=90]{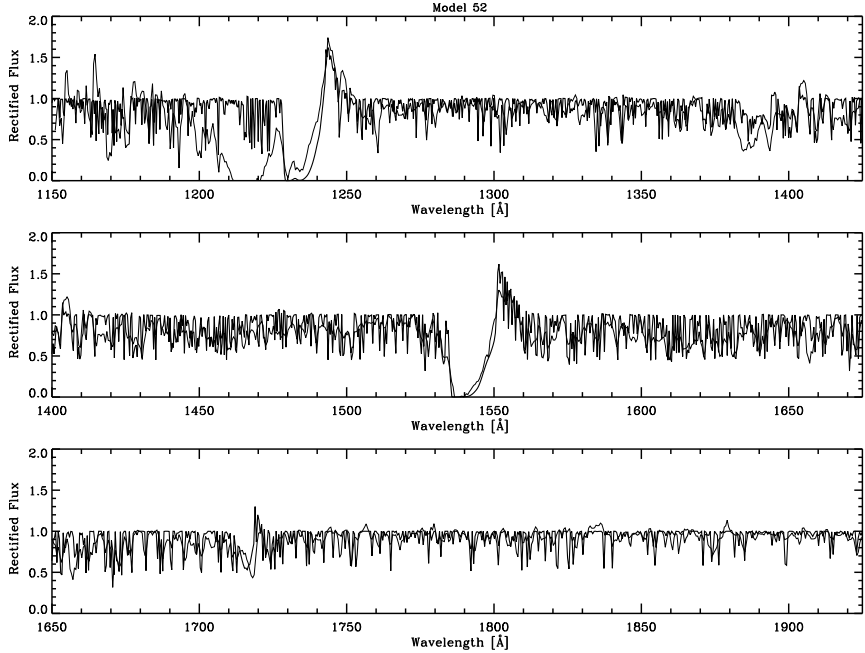} 
 \caption{Comparison of a theoretical stellar UV spectrum with parameters $\log L$ = 5.3, $T_{\rm eff}$ = 34,500 K, $\log g$ = 3.7 (thick line) and 
an average IUE spectrum of three O7.5 III stars (thin). The observed feature at 1216~\AA\ is interstellar Lyman-$\alpha$. From \cite[Leitherer et al. (2010)]{Lei10}.}
   \label{fig2}
\end{center}
\end{figure}

Substantial astrophysical (e.g., inclusion of wind structure and X-ray ionization) and technical (e.g., decrease of computing time by an order of magnitude) improvements
in atmospheric modeling are now allowing us to compute the full spectrum, including the wind lines (\cite[Leitherer et al. 2010]{Lei10}). We used WM-Basic, 
a non-LTE, spherically extended, blanketed, radiation-hydrodynamics code for hot stars (\cite[Pauldrach et al. 2001]{Pau01}) for a self-consistent calculation of the
photospheric and wind parameters and the generation of the synthetic spectrum between 900 and 3000~\AA. The library covers the relevant parameter space of hot stars with
$M \ge 5~M_\odot$ on and
off the main-sequence having abundances between twice and $10^{-2}$ solar. The spectral resolution is 0.5~\AA. An example is shown in Fig.~\ref{fig2} where one particular model is compared to the average
observed IUE spectrum of three stars with closely matching spectral types. The most prominent lines are N~V 1240~\AA, Si~IV 1400~\AA, C~IV 1550~\AA, and N~IV 1720~\AA. Keeping in mind that the slight differences between some features are entirely attributable to mismatches
of stellar parameters, the agreement is excellent and gives confidence in this method.

\begin{figure}[t]
\begin{center}
 \includegraphics[width=4.5in]{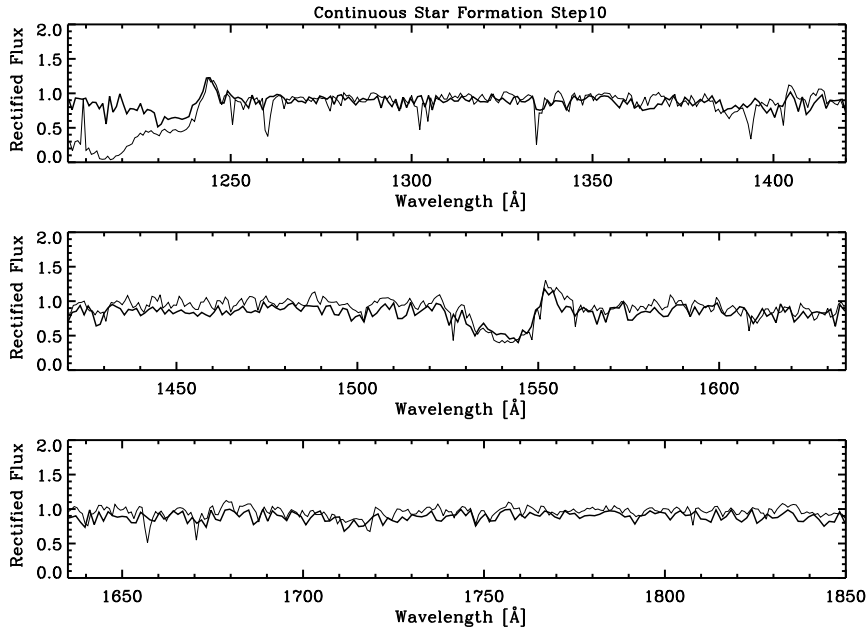} 
 \caption{UV spectrum of a standard stellar population with theoretical (thick line) and empirical (thin) libraries in Starburst99. 
The spectral resolution of the theoretical spectra has been degraded to 1.0~\AA\ to match the empirical spectra. Solar metallicity; Salpeter IMF; age 10 Myr;
continuous star formation. From \cite[Leitherer et al. (2010)]{Lei10}.}
   \label{fig3}
\end{center}
\end{figure}

We then implemented the stellar library in Starburst99 where it complements and extends the empirical libraries of \cite[Robert et al. (1993)]{Rob93}, 
\cite[de Mello et al. (2000)]{Mel00}, \cite[Leitherer et al. (2001)]{Lei01}, and \cite[Pellerin et al. (2002)]{Pel02} currently in use. As a first test we compared 
the synthetic UV spectra for a standard stellar population computed with both methods. This comparison is shown in Fig.~\ref{fig3}. There are no significant differences
that would indicate deficiencies in the models. Several issues are worth noting. (i) The continuum level in the theoretical spectrum is known a priori and no manual normalization
is necessary, whereas such a step had to be performed for the empirical library. The higher continuum level of the empirical spectrum indicates 
a systematic error introduced by the normalization. (ii) N~V 1240~\AA\ in the empirical spectrum is contaminated by Galactic Lyman-$\alpha$, whereas this effect is not
an issue in the theoretical spectrum. (iii) The empirical spectrum is contaminated by narrow interstellar absorption lines which arise along the sight lines to the individual
template stars. Examples are C~II 1335~\AA\ and Si~II 1260~\AA. Even more disturbing is the contamination of the Si~IV 1400~\AA\ stellar-wind line. The theoretical library allows an
analysis without these biases.

\begin{figure}[t]
\begin{center}
 \includegraphics[width=3.8in,angle=90]{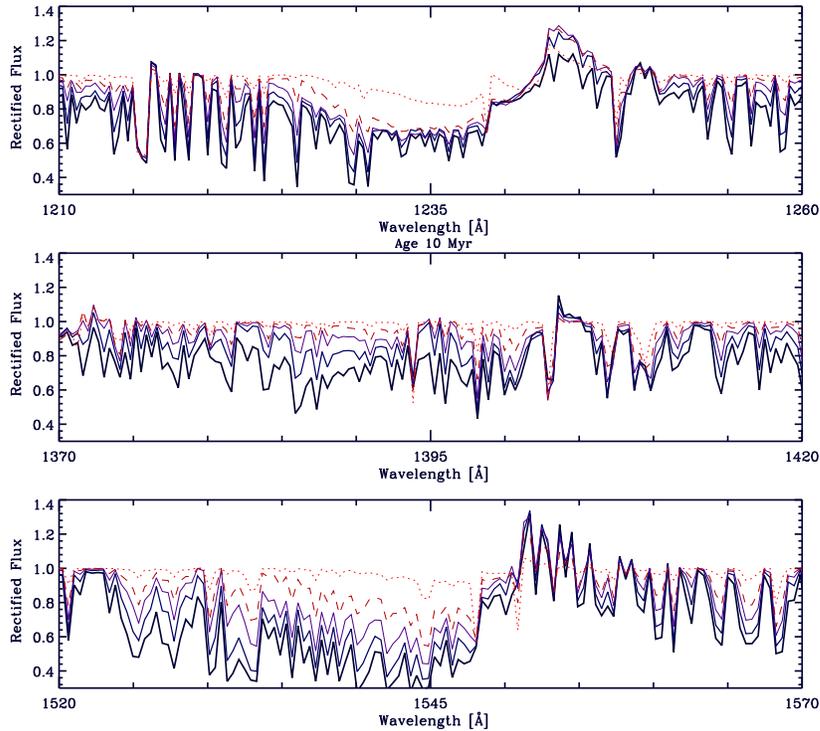} 
 \caption{Variation of N~V 1240~\AA\ (top), Si~IV 1400~\AA\ (middle), or C~IV 1550~\AA\ (bottom) with metallicity. The five metallicities plotted are 2~$Z_\odot$ (thick
solid), $Z_\odot$ (solid), 0.4~$Z_\odot$ (thin solid), 0.2~$Z_\odot$ (dashed), and 0.05~$Z_\odot$ (dotted). The population parameters are as in Fig.~\ref{fig3}. 
From \cite[Leitherer et al. (2010)]{Lei10}.}
   \label{fig4}
\end{center}
\end{figure}

A key application of the newly developed theoretical library will be the determination of stellar abundances in Lyman-break galaxies. Photospheric and stellar-wind
lines respond to abundance variations in a different manner. Whereas photospheric lines are sensitive to excitation/ionization and abundance effects, stellar-wind lines
are in addition predominantly controlled by the stellar mass-loss rate and velocity field. Since the mass-loss rate and velocity field themselves are a function of chemical 
composition, the strong UV lines, such as  N~V 1240~\AA, Si~IV 1400~\AA, or C~IV 1550~\AA\ become quite sensitive to abundance variations despite the fact that they are
deeply saturated. This point is illustrated in Fig.~\ref{fig4}, which shows the variation of N~V, Si~IV, and C~IV with metallicity. All three lines display more
or less significant changes with metallicity, making them suitable as abundance diagnostics. The N~V line, however, is essentially metallicity independent until 
$Z = 0.05~Z_\odot$ is reached.  The behavior of the N V line results from the counteracting effects of chemical composition and stellar temperature on the N$^{4+}$ column density.
The dominant ionization stage in O star winds is N$^{3+}$, and the mean ionization fraction of N$^{4+}$ increases monotonically with stellar temperature. Therefore,
to first order, the N$^{4+}$ column density becomes independent of the stellar mass loss because the ionization fraction and the total nitrogen column density have the opposite
dependence on the mass-loss rate. This effect was discussed before by \cite[Leitherer et al. (2001)]{Lei01}.

The different functional behavior of the various stellar-wind lines allows the discrimination between the often degenerate effects of abundance, age, and IMF. This is of
particular interest at higher and higher redshift and earlier stellar generations, which are expected to have an IMF biased towards very massive stars.

\acknowledgments Support for this work was provided by NASA Grant N1317 and by the STScI Director's Discretionary Research Fund. Part of this research was
done by Paula Ortiz (University of Medell\'{\i}n) as her project during the 2009 STScI Space Astronomy Summer Program.

\

\end{document}